\documentclass[11pt,showpacs]{revtex4}
\def\comment#1{}

\usepackage{graphicx}

\begin{document}

\title{Electron-positron pairs production in a macroscopic charged core}

\author{Remo Ruffini and She-Sheng Xue }

%\email{ruffini@icra.it}\email{    xue@icra.it}

\affiliation{ICRANet Piazzale della Repubblica, 10-65122, Pescara, \\and Physics Department, University of Rome "La Sapienza," P.le A. Moro 5, 00185 Rome, Italy}

%\date{August, 2000}

%\date{\today}

\begin{abstract}
Classical and semi-classical energy states of relativistic electrons bounded by 
a massive and charged core with the charge-mass-radio $Q/M$ 
and macroscopic radius $R_c$ are discussed. We show that the energies of semi-classical (bound) states can be much 
smaller than the negative electron mass-energy ($-mc^2$), and energy-level crossing to negative energy continuum occurs. 
Electron-positron pair production takes place by quantum tunneling, if these bound states are not occupied. 
Electrons fill into these bound states and positrons go to infinity. We explicitly calculate the rate of pair-production, and 
compare it with the rates of electron-positron production by the Sauter-Euler-Heisenberg-Schwinger in a constant electric field. 
In addition, the pair-production rate for the electro-gravitational balance ratio $Q/M = 10^{-19}$ is much larger than the pair-production rate due to the 
Hawking processes. 
\comment{
We point out that in neutral cores with equal proton and electron numbers, 
the configuration of relativistic electrons in these semi-classical (bound) states should be stabilized by photon emissions.
}   
\end{abstract}

%\pacs{12.20ds, 12.20fv, 98.70.S}

\maketitle

\section{Introduction}\label{Z137}

As reviewed in the recent report \cite{report}, very soon after the Dirac equation for a relativistic electron was discovered \cite{z4,z5}, Gordon \cite{z7} (for all $Z< 137$) 
and Darwin \cite{z6} (for $Z=1$) found its solution in the point-like Coulomb potential $V(r)=-Z\alpha/r$,  
they obtained the well-known
Sommerfeld's formula \cite{Sommerfeld} for energy-spectrum, 
\begin{equation}
{\mathcal E}(n,j)=mc^2\left[1+\left(\frac{Z\alpha}{ n-|K|+(K^2-Z^2\alpha^2)^{1/2}}\right)^2\right]^{-1/2},
\label{dirac}
\end{equation}
where the fine-structure constant $\alpha=e^2/\hbar c$, the principle quantum number $n=1,2,3,\cdot\cdot\cdot$ and 
\begin{equation}
K=\left\{\begin{array}{ll} -(j+1/2)= -(l+1), & {\rm if}\quad j=l+\frac{1}{2}, \quad l\ge 0 \\
 (j+1/2)= l, & {\rm if}\quad j=l-\frac{1}{2}, \quad l\ge 1
 \end{array}\right.
\label{dirac-k}
\end{equation}
$l=0,1,2,\cdot\cdot\cdot$ is the orbital angular momentum corresponding to the upper component of Dirac bi-spinor, 
$j$ is the total angular momentum. 
The integer values $n$ and $j$ label bound states whose energies are ${\mathcal E}(n,j)\in (0,mc^2)$. For the example, 
in the case of the lowest energy states, one has      
\begin{eqnarray}
{\mathcal E}(1S_{\frac{1}{2}})&=& mc^2\sqrt {1-(Z\alpha)^2},\label{dirac-k1}\\
{\mathcal E}(2S_{\frac{1}{2}})&=&{\mathcal E}(2P_{\frac{1}{2}})
= mc^2\sqrt{\frac{1+\sqrt{1-(Z\alpha)^2}}{2}},\label{dirac-k2}\\
{\mathcal E}(2P_{\frac{3}{2}})&=& mc^2\sqrt {1-\frac{1}{4}(Z\alpha)^2}.
\label{dirac-k3}
\end{eqnarray}
For all states of the discrete spectrum, the binding energy
$mc^2-{\mathcal E}(n,j)$ increases as the nuclear charge $Z$ increases. 
No regular solution with $n=1,l=0,j=1/2$ and $K=-1$ (the $1S_{1/2}$ ground state)
is found for $Z>137$, this was first noticed by Gordon in his pioneer paper \cite{z7}. 
This is the problem so-called ``$Z=137$ catastrophe''.   

The problem was solved  \cite{g1a,g1b,g1c,g1d,z10,z11,z12,z} by considering the fact that the
nucleus is not point-like and
has an extended charge distribution, and the potential $V(r)$ is not divergent when $r \rightarrow 0$. 
The $Z=137$ catastrophe disappears
and the energy-levels ${\mathcal E}(n,j)$ of the bound states $1S$, $2P$ and $2S$, $\cdot\cdot\cdot$ smoothly continue
to drop toward the negative energy continuum ($E_-<- mc^2$), as $Z$ increases to values larger than $137$.
The critical values $Z_{cr}$ for ${\mathcal E}(n,j)=- mc^2$ were found \cite{z10,z11,z12,z,g1c,rfk78,gbook,krx2008}:
$Z_{cr}\simeq 173$ is a critical value at which the lowest energy-level of the bound state
$1S_{1/2}$ encounters the negative energy continuum, 
while other bound states $2P_{1/2},2S_{3/2},\cdot\cdot\cdot$  
encounter the negative energy continuum at $Z_{cr}>173$, thus energy-level-crossings 
and productions of electron and positron pair takes place, provided these bound states are unoccupied.  
We refer the readers to \cite{z10,z11,z12,z,popov1972,popov2001,rfk78,gbook,krx2008} for mathematical and numerical details.

The energetics of this phenomenon can be understood as follow.
The energy-level of the bound state $1S_{1/2}$ can be estimated as follow,
\begin{equation}
{\mathcal E}(1S_{1/2})=mc^2 - \frac{Z e^2}{\bar r}<-mc^2,
\label{1S}
\end{equation}
where $\bar r $ is the average radius of the $1S_{1/2}$ state's orbit, and the binding energy of this
state $Ze^2/\bar r > 2 mc^2$. If this bound state is unoccupied, 
the bare nucleus gains a binding energy $Ze^2/\bar r$ larger than 
$2mc^2$, and becomes unstable against the production of an electron-positron pair. Assuming this 
pair-production occur around the radius $\bar r$, we have energies of electron ($\epsilon_-$) and positron ($\epsilon_+$):
\begin{equation}
\epsilon_-=\sqrt{(c|{\bf p}_-|)^2+m^2c^4}-\frac{Z e^2}{\bar r};\ \quad \epsilon_+=\sqrt{(c|{\bf p}_+|)^2+m^2c^4}+\frac{Z e^2}{\bar r},
\label{eofep}
\end{equation}
where ${\bf p}_\pm$ are electron and positron momenta, and ${\bf p}_-=-{\bf p}_+$. 
The total energy required for a pair production is,
\begin{equation}
\epsilon_{-+}=\epsilon_-+\epsilon_+=2\sqrt{(c|{\bf p}_-|)^2+m^2c^4},
\label{totaleofep}
\end{equation}
which is independent of the potential $V(\bar r)$. The potential energies $\pm eV(\bar r)$ of electron 
and positron cancel 
each other and do not contribute to the total energy (\ref{totaleofep}) required for pair production. 
This energy (\ref{totaleofep}) is acquired from the binding energy ($Ze^2/\bar r > 2 mc^2$)
by the electron filling into the bound state $1S_{1/2}$. A part of the binding energy becomes 
the kinetic energy of positron that goes out.  
This is analogous to the familiar case that a proton ($Z=1$)
catches an electron into the ground state $1S_{1/2}$, and a photon is emitted with the energy not less than
13.6 eV.   
  
In this article, we study classical and semi-classical states of electrons,   
electron-positron pair production 
in an electric potential of macroscopic cores with charge $Q=Z|e|$, mass $M$ and macroscopic radius $R_c$.   

\section{Classical description of electrons in potential of cores}\label{class}

\subsection{\it Effective potentials for particle's radial motion}\label{eff}

Setting the origin of spherical coordinates $(r,\theta,\phi)$ at the center of such cores, we write the 
vectorial potential $A_\mu=({\bf A}, A_0)$, where ${\bf A}=0$ and $A_0$ is the Coulomb potential. The motion 
of a relativistic electron with mass $m$ and charge $e$ is described by its radial momentum $p_r$, total angular momenta 
$p_\phi$ and the Hamiltonian,
\begin{eqnarray}
H_\pm &=& \pm mc^2\sqrt{1+(\frac{p_r}{mc})^2+(\frac{p_\phi}{mcr})^2}-V(r),
\label{toth}
\end{eqnarray}
where the potential energy $V(r)=eA_0$, and $\pm$ corresponds for positive and negative energies.
The states corresponding to negative energy solutions are fully occupied. 
The total angular momentum $p_\phi$ is conserved, for the potential $V(r)$ is spherically symmetric. 
For a given angular momentum $p_\phi=mv_\perp r$, where $v_\perp$ is the transverse velocity, the effective potential energy for electron's radial motion is 
\begin{eqnarray}
E_\pm(r) &=& \pm mc^2\sqrt{1+(\frac{p_\phi}{mcr})^2}-V(r),
\label{tote}
\end{eqnarray} 
where  $\pm$ indicates positive and negative effective energies, outside the core ($r\geq R_c$), the Coulomb potential energy $V(r)$ is given by
\begin{equation}
V_{\rm out}(r)=\frac{Ze^2}{r}.
\label{potenoutside}
\end{equation}
Inside the core ($r\leq R_c$), the Coulomb potential energy is given by
\begin{equation}
V_{\rm in}(r) = \frac{Ze^2}{2R_c} \left [3-\left(\frac{r}{R_c}\right)^2\right],
\label{poteninside}
\end{equation}
where we postulate the charged core has a uniform charge distribution with constant charge density $\rho=Ze/V_c$, and the core 
volume $V_c=4\pi R^3_c/3$. Coulomb potential energies outside the core (\ref{potenoutside}) and 
inside the core (\ref{poteninside}) are continuous at $r=R_c$. 
The electric field on the surface of the core, 
\begin{equation}
E_s=\frac{Q}{R_c^2}=\beta\frac{\lambda_e}{R_c}E_c,\quad \beta\equiv \frac{Ze^2}{mc^2R_c}
\label{mtote6}
\end{equation}
where the electron Compton wavelength $\lambda_e=\hbar/(mc)$, the critical electric field $E_c=m^2c^3/(e\hbar)$ and
the parameter $\beta$ is the electric potential-energy on the surface of the core in unit 
of the electron mass-energy.

\subsection{\it Stable classical orbits (states) outside the core.}\label{outside}

\begin{figure}[th] 
\begin{center}
\includegraphics[width=\hsize,clip]{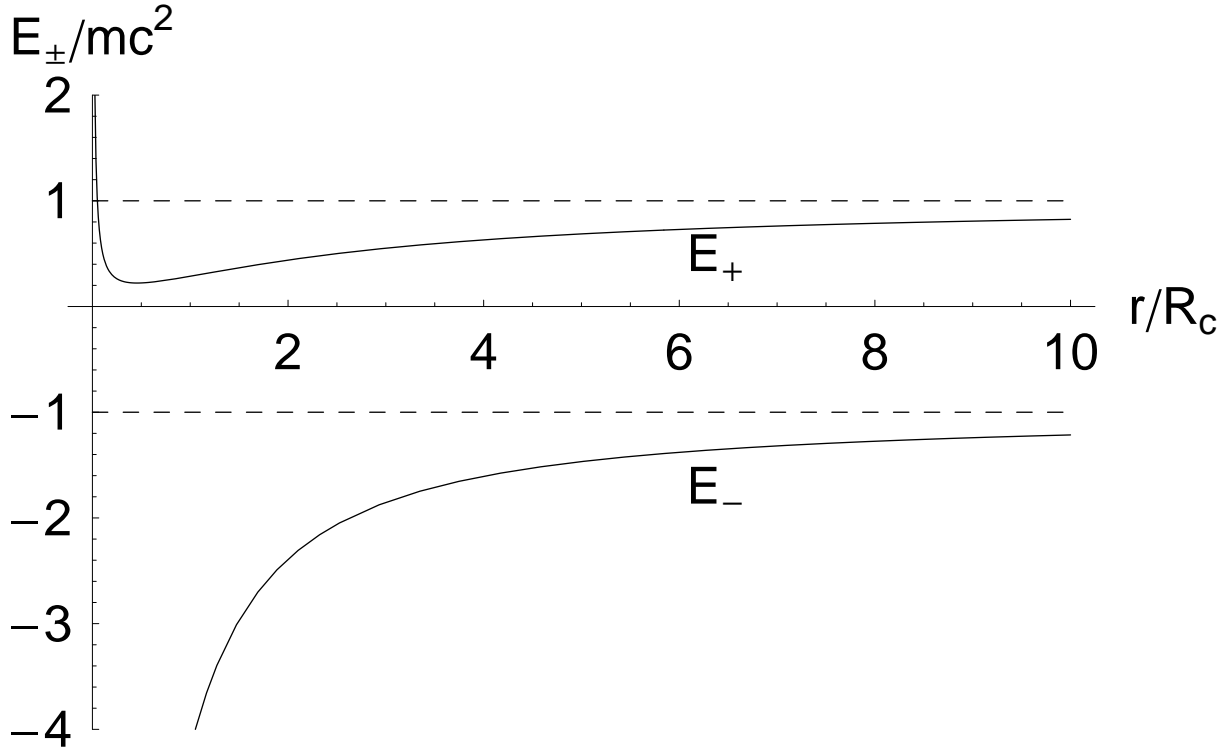}
\end{center}
\caption{In the case of point-like charge distribution, 
we plot the positive and negative effective potential energies $E_\pm$ (\ref{tote}), $p_\phi/(mcR_c)=2$ and $Ze^2=1.95mc^2R_c$,
to illustrate the radial location $R_L$ (\ref{stable})
 of stable orbits where $E_+$ has a minimum (\ref{mtote}). All stable orbits are 
described by $cp_\phi> Ze^2$. The last stable orbits are given by $cp_\phi\rightarrow Ze^2+0^+$, 
whose radial location $R_L\rightarrow 0$ and energy ${\mathcal E}\rightarrow 0^+$.
There is no any stable orbit with energy ${\mathcal E}< 0$ and the energy-level crossing with 
the negative energy spectrum $E_-$ is impossible.}%
\label{fig1}%
\end{figure}

Given different values of total angular momenta $p_\phi$, the stable circulating orbits $R_L$ (states) 
are determined by the minimum of the effective potential $E_+(r)$ (\ref{tote}) (see Fig.~\ref{fig1}), at which $dE_+(r)/dr =0$.
We obtain stable orbits locate at the radii $R_L$ outside the core,
\begin{equation}
R_L=\left(\frac{p_\phi^2}{Ze^2m}\right)\sqrt{1-\left(\frac{Ze^2}{cp_\phi}\right)^2},\quad R_L\ge R_c,
\label{stable}
\end{equation}
for different $p_\phi$-values.
Substituting Eq.~(\ref{stable}) into Eq.~(\ref{tote}), we find the energy of electron at each stable orbit,
\begin{equation}
{\mathcal E}\equiv {\rm min}(E_+) = mc^2\sqrt{1-\left(\frac{Ze^2}{cp_\phi}\right)^2}.
\label{mtote}
\end{equation}
For the condition $R_L\gtrsim R_c$, we have 
\begin{equation}
\left(\frac{Ze^2}{cp_\phi}\right)^2 \lesssim \frac{1}{2} \left[\beta(4+\beta^2)^{1/2}-\beta^2\right],
\label{ocon}
\end{equation}
where the semi-equality holds for the last stable orbits outside the core $R_L\rightarrow R_c+0^+$. In the point-like case $R_c\rightarrow 0$,
the last stable orbits are  
\begin{equation}
cp_\phi\rightarrow Ze^2+0^+,\quad R_L\rightarrow 0^+,\quad {\mathcal E} \rightarrow 0^+.
\label{ocon1}
\end{equation}
Eq.~(\ref{mtote}) shows that there are only
positive or null energy solutions (states) in the case of 
a point-like charge, which corresponds to the energy-spectra 
%(\ref{dirac}) (see 
Eqs.~(\ref{dirac-k1},\ref{dirac-k2},\ref{dirac-k3}) 
in quantum mechanic scenario. 
While for $p_\phi\gg 1$, radii of stable orbits $R_L\gg 1$ and energies ${\mathcal E}\rightarrow mc^2+0^-$, 
classical electrons in these orbits are critically bound for their banding energy goes to zero.   
We conclude that the energies (\ref{mtote}) of stable orbits outside the core 
must be smaller than $mc^2$, but larger than zero, 
${\mathcal E}> 0$. Therefore, no energy-level crossing with the 
negative energy spectrum occurs. 
    
\subsection{\it Stable classical orbits inside the core.}\label{inside}

We turn to the stable orbits of electrons 
inside the core. Analogously, using Eqs.~(\ref{tote},\ref{poteninside}) 
and  $dE_+(r)/dr =0$, we obtain the stable orbit radius $R_L\leq 1$ in the unit of $R_c$, obeying the following equation,
\begin{equation}
\beta^2 (R_L^8+ \kappa^2R_L^6) = \kappa^4;\quad \kappa=\frac{p_\phi}{mcR_c}.
\label{inorbite}
\end{equation}
and corresponding to the minimal energy (binding energy) 
of these states
\begin{equation}
{\mathcal E} = \frac{Ze^2}{R_c}\left[\left(\frac{cp_\phi}{Ze^2}\right)^2\frac{1}{R_L^4}-\frac{1}{2}(3-R_L^2)\right].
\label{mtote2}
\end{equation}
There are 8 solutions to this polynomial equation (\ref{inorbite}), only one is physical,  the solution $R_L$ that has to be
real, positive and smaller than one. As example, the numerical solution to Eq.~(\ref{inorbite}) is $R_L=0.793701$ 
for $\beta=4.4\cdot 10^{16}$ and $\kappa=2.2\cdot 10^{16}$. In following, we 
respectively adopt non-relativistic and ultra-relativistic approximations
to obtain analytical solutions.

First considering the non-relativistic case for those stable orbit
states whose kinetic energy term
characterized by angular momentum term $p_\phi$, see 
Eq.~(\ref{tote}), is much smaller than the 
rest mass term $mc^2$, we obtain the following approximate equation,
\begin{equation}
\beta^2 R_L^8 \simeq \kappa^4,
\label{inorbite3}
\end{equation} 
and the solutions for stable orbit radii are,
\begin{equation}
R_L \simeq \frac{\kappa^{1/2}}{\beta^{1/4}}=\left(\frac{cp_\phi}{Ze^2}\right)^{1/2}\beta^{1/4}<1,
\label{inorbite4}
\end{equation}
and energies,
\begin{equation}
{\mathcal E} \simeq \left(1-\frac{3}{2}\beta+\frac{1}{2}\kappa\beta^{1/2}\right)mc^2.
\label{mtote3n}
\end{equation}
The consistent conditions for this solution are $\beta^{1/2}>\kappa$ for $R_L<1$, and $\beta \ll 1$ 
for non-relativistic limit $v_\perp\ll c$, where the transverse velocity $v_\perp=p_\phi/(mR_L)$.
As a result, the binding energies (\ref{mtote3n}) of these states are $mc^2>{\mathcal E}>0$, are never less than zero. These in fact correspond to the stable states which have large radii closing to the radius $R_c$ of cores
and $v_\perp \ll c$. 

Second considering the ultra-relativistic case for those stable orbit
states whose the kinetic energy term
characterized by angular momentum term $p_\phi$, see 
Eq.~(\ref{tote}), is much larger than the rest mass term $mc^2$, 
we obtain the following approximate equation, 
\begin{equation}
\beta^2 R_L^6 \simeq \kappa^2,
\label{inorbite1}
\end{equation} 
and the solutions for stable orbit radii are,
\begin{equation}
R_L \simeq \left(\frac{\kappa}{\beta}\right)^{1/3}=\left(\frac{p_\phi c}{Ze^2}\right)^{1/3}<1,
\label{inorbite2}
\end{equation}
which gives $R_L\simeq 0.7937007$ for the same values of parameters $\beta$ and $\kappa$ in above. 
The consistent condition for this solution is $\beta> \kappa\gg 1$ for $R_L<1$. The energy levels of these ultra-relativistic states
are,
\begin{equation}
{\mathcal E} \simeq \frac{3}{2}\beta\left[\left(\frac{p_\phi c}{Ze^2}\right)^{2/3}-1\right]mc^2,
\label{mtote3}
\end{equation}
and $mc^2 > {\mathcal E}>-1.5\beta mc^2$. The particular  
solutions ${\mathcal E}=0$ and ${\mathcal E}\simeq -mc^2$ are respectively given by 
\begin{eqnarray}
\left(\frac{p_\phi c}{Ze^2}\right)\simeq 1;\quad
\left(\frac{p_\phi c}{Ze^2}\right)\simeq \left(1-\frac{2}{3\beta}\right)^{3/2}\label{inneg}.
\end{eqnarray}
These in fact correspond to the stable states which have small radii closing to the center of cores and $v_\perp \lesssim c$. 

To have the energy-level crossing to the negative energy continuum, we are interested in the values $\beta>\kappa\gg 1$ for which 
the energy-levels (\ref{mtote3}) of stable orbit states are equal to or 
less than $-mc^2$,
\begin{equation}
{\mathcal E}\simeq\frac{3}{2}\beta\left[\left(\frac{p_\phi c}{Ze^2}\right)^{2/3}-1\right]mc^2\leq -mc^2.
\label{mtote4}
\end{equation} 
As example, with $\beta =10$ and $\kappa =2$, $R_L\simeq 0.585$, ${\mathcal E}_{\rm min}\simeq -9.87 mc^2$.
The lowest energy-level of electron state is $p_\phi /(Ze^2)=\kappa/\beta\rightarrow 0$ with the binding energy,
\begin{equation}
{\mathcal E}_{\rm min} = -\frac{3}{2}\beta mc^2,
\label{mtote5}
\end{equation}
locating at $R_L\simeq (p_\phi c/Ze^2)^{1/3}\rightarrow 0$, the bottom of the potential 
energy $V_{\rm in}(0)$ (\ref{poteninside}). 

\section{\it Semi-Classical description}\label{semi}

\subsection{Bohr-Sommerfeld quantization}

In order to have further understanding, we consider the semi-classical scenario. Introducing the Planck constant $\hbar=h/(2\pi)$, we adopt
the semi-classical Bohr-Sommerfeld quantization rule
\begin{equation}
\int p_\phi d\phi \simeq h (l+\frac{1}{2}),\quad \Rightarrow\quad
p_\phi(l) \simeq \hbar(l+\frac{1}{2}),\quad l=0,1,2,3,\cdot\cdot\cdot,
\label{angq}
\end{equation}
which are discrete values selected from continuous total angular momentum $p_\phi$ in the classical scenario. The variation of total 
angular momentum $\Delta p_\phi = \pm \hbar $ in th unit of the Planck constant $\hbar$, we make 
substitution 
\begin{eqnarray}
\left(\frac{p_\phi c}{Ze^2}\right)\Rightarrow \left(\frac{2l+1}{2Z\alpha}\right),\quad \alpha=\frac{e^2}{(\hbar c)}
\label{qsub}, 
\end{eqnarray}  
in classical solutions that we obtained in section (\ref{class}).

\begin{enumerate}

\item The radii and energies of stable states outside the core (\ref{stable}) and (\ref{mtote}) become:
\begin{eqnarray}
R_L &=& \lambda\left(\frac{2l+1}{Z\alpha}\right)\sqrt{1-\left(\frac{2Z\alpha}{2l+1}\right)^2},
\label{qstable}\\
{\mathcal E} & = & mc^2\sqrt{1-\left(\frac{2Z\alpha}{2l+1}\right)^2}, \label{qmtote}
\end{eqnarray}
where the electron Compton length $\lambda=\hbar/(mc)$.

\item The radii and energies of non-relativistic stable states inside the core (\ref{inorbite4}) and (\ref{mtote3n}) become:
\begin{eqnarray}
R_L &\simeq & \left(\frac{2l+1}{2Z\alpha}\right)^{1/2}\beta^{1/4},
\label{qinorbite4}\\
{\mathcal E} &\simeq & \left(1-\frac{3}{2}\beta+\frac{\lambda (2l+1)}{4R_c}\beta^{1/2}\right)mc^2.
\label{qmtote3n}
\end{eqnarray}

\item The radii and energies of ultra-relativistic stable states inside the core (\ref{inorbite2}) and (\ref{mtote3}) become:
\begin{eqnarray}
R_L & \simeq & \left(\frac{2l+1}{2Z\alpha}\right)^{1/3},
\label{qinorbite2}\\
{\mathcal E} &\simeq & \frac{3}{2}\beta\left[\left(\frac{2l+1}{2Z\alpha}\right)^{2/3}-1\right]mc^2.
\label{qmtote3}
\end{eqnarray}

\end{enumerate}
Note that radii $R_L$ in the second and third cases are in unit of $R_c$.  

\subsection{Stability of semi-classical states}

When these semi-classical states are not occupied as required by the Pauli Principle, 
the transition from one state 
to another with different discrete values of total angular momentum $l$ ($l_1,l_2$ and $\Delta l=l_2-l_1=\pm 1$) 
undergoes by emission or absorption of a spin-1 ($\hbar$) 
photon. Following the energy and angular-momentum conservations, photon emitted or absorbed in the transition have angular momenta 
$p_\gamma=p_\phi(l_2)-p_\phi(l_1)=\hbar (l_2-l_1)=\pm\hbar$ and energy 
${\mathcal E}_\gamma={\mathcal E}(l_2)-{\mathcal E}(l_1)$. In this transition of stable states, the variation of radius is $\Delta R_L=R_L(l_2)-R_L(l_1)$.

We first consider the stability of semi-classical states against such transition in the case of point-like charge, 
i.e., Eqs.~(\ref{qstable},\ref{qmtote}) with $l=0,1,2,\cdot\cdot\cdot$.  
As required by the Heisenberg indeterminacy principle 
$\Delta\phi\Delta p_\phi \simeq 4\pi p_\phi(l) \gtrsim h$,
the absolute ground state for minimal energy and angular momentum is given by the $l=0$ state, $p_\phi\sim \hbar/2$, $R_L\sim  \lambda (Z\alpha)^{-1}(1-(2Z\alpha)^2)^{1/2}>0$
and ${\mathcal E} \sim  mc^2(1-(2Z\alpha)^2)^{1/2}>0$, which corresponds to the last stable orbit (\ref{ocon1}) in the classical scenario.
Thus the stability of all semi-classical states $l>0$ is guaranteed by the Pauli principle. This is only case for $Z\alpha \le 1/2$. 
While for $Z\alpha > 1/2$, there is not an absolute ground state
in the semi-classical scenario. This can be understood by  
examining how the lowest energy states are selected by the quantization rule in the semi-classical scenario
out of the last stable orbits (\ref{ocon1}) in the classical scenario. For the case of $Z\alpha\le 1/2$, equating $p_\phi$ in 
Eq.~(\ref{ocon1}) to $p_\phi=\hbar(l+ 1/2)$ (\ref{angq}), we find the selected state $l=0$ is only possible solution 
so that the ground state $l=0$ in the semi-classical scenario corresponds to the last stable orbits (\ref{ocon1}) 
in the classical scenario. While for the case of $Z\alpha > 1/2$, equating $p_\phi$ in Eq.~(\ref{ocon1}) to
$p_\phi=\hbar(l+ 1/2)$ (\ref{angq}), we find the selected semi-classical state 
\begin{equation}
\tilde l= \frac{Z\alpha-1}{2}>0,
\label{is0}
\end{equation} 
in the semi-classical scenario corresponds to the last stable orbits (\ref{ocon1}) in the classical scenario. This state $l=\tilde l>0$
is not protected by the Heisenberg indeterminacy principle from quantum-mechanically decaying in $\hbar$-steps to the states 
with lower angular momenta and energies (correspondingly smaller radius $R_L$ (\ref{qstable})) via photon emissions. This clearly shows
that the ``$Z=137$-catastrophe'' corresponds to $R_L\rightarrow 0$, falling to the center of the Coulomb potential 
and all semi-classical states ($l$) are unstable.

Then we consider the stability of semi-classical states against such transition in the case of charged cores $R_c\not=0$.
Substituting $p_\phi$ in Eq.~(\ref{angq}) into Eq.~(\ref{ocon}), we obtain the selected 
semi-classical state $\tilde l$ corresponding to the 
last classical stable orbit outside the core,
\begin{equation}
\tilde l= \sqrt{2}\left(\frac{R_c}{\lambda}\right)\left[\left(\frac{4R_c}{Z\alpha\lambda}+1\right)^{1/2}-1\right]^{-1/2}\approx 
(Z\alpha)^{1/4}\left(\frac{R_c}{\lambda}\right)^{3/4}> 0.
\label{is1}
\end{equation}
Analogously to Eq.~(\ref{is0}), the same argument concludes the instability of this semi-classical state, which must quantum-mechanically 
decay to states with angular momentum $l<\tilde l$ inside the core, provided these semi-classical states are not occupied. This conclusion is 
independent of $Z\alpha$-value.

We go on to examine the stability of semi-classical states inside the core. In the non-relativistic case $(1\gg \beta >\kappa^2)$, the last classical 
stable orbits locate at $R_L\rightarrow 0$ and $p_\phi\rightarrow 0$ given by Eqs.~(\ref{inorbite4},\ref{mtote3n}), 
corresponding to the lowest semi-classical 
state (\ref{qinorbite4},\ref{qmtote3n}) with $l=0$ and energy $mc^2>{\mathcal E}>0$. 
In the ultra-relativistic case $(\beta >\kappa \gg 1)$, the last classical 
stable orbits locate at $R_L\rightarrow 0$ and $p_\phi\rightarrow 0$ given by Eqs.~(\ref{inorbite2},\ref{mtote3}), 
corresponding to the lowest semi-classical 
state (\ref{qinorbite2},\ref{qmtote3}) with $l=0$ and minimal energy,
\begin{eqnarray}
{\mathcal E} \simeq  \frac{3}{2}\beta\left[\left(\frac{1}{2Z\alpha}\right)^{2/3}-1\right]mc^2\approx - \frac{3}{2}\beta mc^2.
\label{qmin}
\end{eqnarray}
This concludes that the $l=0$ semi-classical state inside the core is an absolute ground state in both non- and ultra-relativistic cases. 
The Pauli principle assures that all semi-classical states $l>0$ are stable, provided all these states accommodate electrons. The electrons
can be either present inside the core or produced from the vacuum polarization, later will be discussed in details.

We are particular interested in the ultra-relativistic case $\beta > \kappa \gg 1$, i.e., $Z\alpha \gg 1$, the energy-levels 
of semi-classical states can be more profound than $-mc^2$ (${\mathcal E}< -mc^2$), energy-level crossings and pair-productions occur 
if these states are unoccupied, as discussed in introductory section. 
\comment{
It is even more important to mention that neutral cores 
like neutron stars of proton number $Z\sim 10^{52}$, the Thomas-Fermi approach has to be adopted to find 
the configuration of electrons in these semi-classical states, which has the depth of energy-levels 
${\mathcal E}\sim -m_\pi c^2$ to accommodate electrons and a supercritical electric field ($E>E_c$) 
on the surface of the core \cite{rrx2007,rrx2008}.    
}
      
\section{\it Production of electron-positron pair }\label{tunneling}

\begin{figure}[th] 
\begin{center}
\includegraphics[width=\hsize,clip]{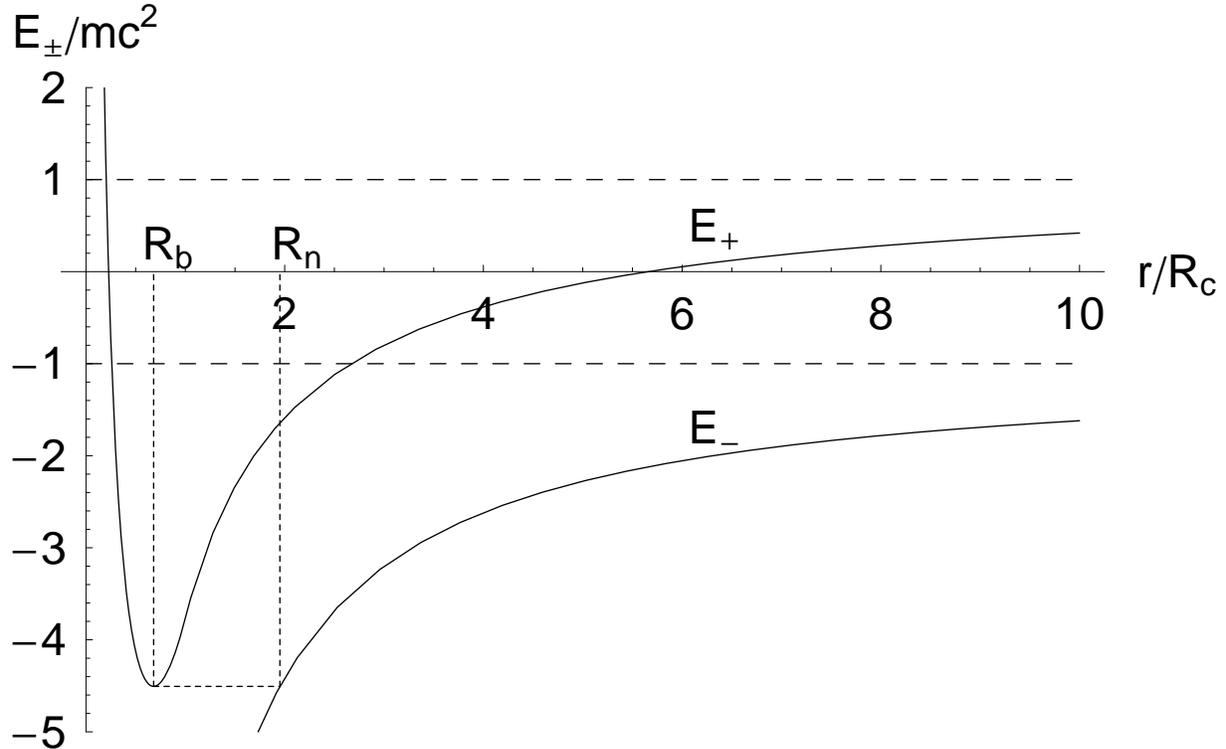}
\end{center}
\caption{For the core $\kappa=2$ and $\beta=6$, 
we plot the positive and negative effective potentials $E_\pm$ (\ref{tote}) ,
in order to illustrate the radial location (\ref{inorbite2})
$R_L<R_c$ of stable orbit, where $E_+$'s minimum (\ref{mtote3}) ${\mathcal E}<mc^2$ is. All stable orbits inside the core
are described by $\beta>\kappa>1$. The last stable orbit is given by $\kappa/\beta\rightarrow 0$, 
whose radial location $R_L\rightarrow 0$ and energy ${\mathcal E}\rightarrow {\mathcal E}_{\rm min}$ (\ref{mtote5}).
We indicate that the energy-level crossing between bound state (stable orbit) energy at $R_L=R_b$ 
and negative energy spectrum 
$E_-$ (\ref{mtote3}) at the turning point $R_n$. }%
\label{fig2}%
\end{figure}

When the energy-levels of semi-classical (bound) states ${\mathcal E} \leq -mc^2$ (\ref{mtote4}), 
energy-level crossings between these energy-levels (\ref{mtote3}) and negative energy continuum (\ref{tote}) for $p_r=0$,
as shown in Fig.~\ref{fig2}. 
The energy-level-crossing indicates that 
${\mathcal E}$ (\ref{mtote3}) and $E_-$ (\ref{tote}) 
are equal,
\begin{equation}
{\mathcal E}=E_-,
\label{energylcrossing}
\end{equation}
where angular momenta $p_\phi$ in ${\mathcal E}$ (\ref{qmtote3}) and $E_-$ (\ref{tote}) 
are the same for angular-momentum conservation. 
The production of electron-positron pairs must takes place, provided these semi-classical (bound) states are unoccupied. 
The phenomenon of pair production can be understood as a
quantum-mechanical tunneling process of relativistic electrons.
The energy-levels ${\mathcal E}$ of semi-classical (bound) states are given by Eq.~(\ref{qmtote3}) or (\ref{mtote4}). 
The probability amplitude for this
process can be calculated by a semi-classical WKB method \cite{krx2008}:
\begin{eqnarray}
W_{\rm WKB}(|{\bf p}_\perp |) &\equiv  & \exp\left\{-\frac{2}{\hbar}\int_{R_b}^{R_n} p_rdr\right\},
\label{tprobability1}
\end{eqnarray}
where $|{\bf p}_\perp|=p_\phi/r$ is transverse momenta and the radial momentum,
\begin{eqnarray}
p_r(r) &=& \sqrt{(c|{\bf p}_\perp|) ^2+m^2c^4-[{\mathcal E}+V(r)]^2}.
\label{px}
\end{eqnarray}
The energy potential $V(r)$ is either given by $V_{\rm out}(r)$ (\ref{potenoutside}) for $r>R_c$, or 
$V_{\rm in}(r)$ (\ref{poteninside}) for $r<R_c$.
The limits of integration (\ref{tprobability1}):
$R_b=R_L < R_c$ (\ref{inorbite2}) or (\ref{qinorbite2}) indicating 
the location of the classical orbit (classical turning point) of semi-classical (bound) state; 
while another classical turning point $R_n$ 
is determined by setting $p_r(r)=0$ in Eq.~(\ref{px}). 
There are two cases: $R_n<R_c$ and $R_n>R_c$, depending on $\beta$ and $\kappa$ values. 

To obtain a maximal WKB-probability amplitude (\ref{tprobability1}) of pair production, we only consider 
the case that the charge core is bare and 
\begin{itemize}

\item the lowest energy-levels of 
semi-classical (bound) states: $p_\phi/(Ze^2)=\kappa/\beta\rightarrow 0$, the location of classical orbit (\ref{inorbite2}) 
$R_L=R_b\rightarrow 0$ and energy (\ref{mtote3}) ${\mathcal E}\rightarrow {\mathcal E}_{\rm min}=-3\beta mc^2/2$ (\ref{mtote5});

\item another classical turning point $R_n\le R_c$, since the probability is exponentially suppressed 
by a large tunneling length $\Delta=R_n-R_b$. 

\end{itemize}
In this case ($R_n\le R_c$), Eq.~(\ref{px}) becomes
\begin{eqnarray}
p_r &=&\sqrt{(c|{\bf p}_\perp |)^2+ m^2c^4}\sqrt{1-\frac{\beta^2m^2c^4}{4[(c|{\bf p}_\perp |)^2+ m^2c^4]}\left(\frac{r}{R_c}\right)^4},
\label{px1}
\end{eqnarray}
and $p_r=0$ leads to
\begin{eqnarray}
\frac{R_n}{R_c}=\left(\frac{2}{\beta mc^2}\right)^{1/2}[(c|{\bf p}_\perp |)^2+ m^2c^4]^{1/4}.
\label{rn1}
\end{eqnarray}
Using Eqs.~(\ref{tprobability1},\ref{px1},\ref{rn1}), we have
\begin{eqnarray}
W_{\rm WKB}(|{\bf p}_\perp |) 
&=& \exp\left\{-{\frac{2^{3/2}[(c|{\bf p}_\perp |)^2+ m^2c^4]^{3/4}R_c}{c\hbar(mc^2\beta)^{1/2}}}\int_{0}^1 \sqrt{1-x^4}dx\right\}\nonumber\\
&= &\exp\left\{-0.87{\frac{2^{3/2}[(c|{\bf p}_\perp |)^2+ m^2c^4]^{3/4}R_c}{c\hbar(mc^2\beta)^{1/2}}}\right\}.
\label{tprobability3}
\end{eqnarray}
Dividing this probability amplitude 
by the tunneling length $\Delta\simeq R_n$ and time interval $\Delta t\simeq  2\pi\hbar/(2mc^2)$ 
in which the quantum tunneling occurs, and integrating over two spin states and the transverse phase-space 
$2\int d{\bf r}_\perp d{\bf p}_\perp /(2\pi\hbar)^2$, we approximately obtain the rate of pair-production per the unit of time and volume,
\begin{eqnarray}
\Gamma_{\rm NS}\equiv \frac{d^4N}{dtd^3x}
&\simeq &  \frac{1.15}{6\pi^2}\left(\frac{Z\alpha}{\tau R_c^3}\right)
\exp\left\{-{\frac{2.46 }{(Z\alpha)^{1/2}}}\left(\frac{R_c}{\lambda}\right)^{3/2}\right\},
\label{rate1}\\
&=&  \frac{1.15}{6\pi^2}\left(\frac{\beta}{\tau\lambda R_c^2}\right)\exp\left\{-\frac{2.46 R_c}{\beta^{1/2}\lambda}\right\},
\label{rate2}\\
&= &  \frac{1.15}{6\pi^2}\left(\frac{1}{\tau\lambda^2R_c}\right)\left(\frac{E_s}{E_c}\right)
\exp\left\{-2.46\left(\frac{R_c}{\lambda}\right)^{1/2}\left(\frac{E_c}{E_s}\right)^{1/2}\right\},
\label{rate3}
\end{eqnarray}
where $E_s=Ze/R_c^2$ is the electric field on the surface of the core and the Compton time $\tau=\hbar/mc^2$. 
\comment{
To have the size of this pair-production rate, we compare it with the Sauter-Euler-Heisenberg-Schwinger rate 
of pair-production in a constant field $E$ \cite{sauter,euler,schwinger},
\begin{eqnarray}
\Gamma_{\rm S}\equiv \frac{d^4N}{dtd^3x}
&\simeq &  \frac{1}{4\pi^3\tau\lambda^3}\left(\frac{E}{E_c}\right)^2\exp\left\{-\pi\frac{E_c}{E}\right\}.
\label{rates}
\end{eqnarray}
When the parameter $\beta\simeq (R_c/\lambda)^2$, Eq.~(\ref{rate2}) becomes
\begin{eqnarray}
\Gamma_{\rm NS}\equiv \frac{d^4N}{dtd^3x}
\simeq \frac{1.15}{6\pi^2}\left(\frac{1}{\tau\lambda^3}\right)\exp\left\{-2.46 \right\}=1.66\cdot 10^{-3}/(\tau\lambda^3),
\label{rate21}
\end{eqnarray}
which is close to the Sauter-Euler-Heisenberg-Schwinger rate (\ref{rates}) $\Gamma_{\rm S}\simeq 3.5\cdot 10^{-4}/(\tau\lambda^3)$ 
at $E\simeq E_c$. 
we have $R_c/\lambda=2.59\cdot 10^{16}$ 
and $\beta= 3.86\cdot 10^{-17}Z\alpha$, leading to $Z\simeq 2.4\cdot 10^{51}$ and the electric field on the core surface 
$E_s/E_c=Z\alpha(\lambda/R_c)^2\simeq 2.6 \cdot 10^{16}$. In this case, the charge-mass radio 
$Q/(G^{1/2}M)=2\cdot 10^{-6}|e|/(G^{1/2}m_p)=2.2\cdot 10^{12}$, where where $G$ is the Newton constant and 
proton's charge-mass radio $|e|/(G^{1/2}m_p)=1.1\cdot 10^{18}$.
}

To have the size of this pair-production rate, we consider a macroscopic core of mass $M=M_\odot$ and radius $R_c=10$km, and the electric field on the core 
surface $E_s$ (\ref{mtote6}) is about the critical field ($E_s\simeq E_c$). In this case, 
$Z=\alpha^{-1}(R_c/\lambda)^2\simeq 9.2\cdot 10^{34}$, 
$\beta =Z\alpha \lambda/R_c=R_c/\lambda \simeq 2.59\cdot 10^{16}$,
and the rate (\ref{rate2}) becomes
\begin{eqnarray}
\Gamma_{\rm NS}\equiv \frac{d^4N}{dtd^3x}
\simeq \frac{1.15}{6\pi^2}\left(\frac{1}{\tau\lambda^3}\right)\left(\frac{\lambda}{R_c}\right)
\exp\left\{-2.46 \left(\frac{R_c}{\lambda}\right)\right\},
\label{rate22}
\end{eqnarray}
which is exponentially small for $R_c\gg \lambda$. In this case, the charge-mass radio 
$Q/(G^{1/2}M)=2\cdot 10^{-6}|e|/(G^{1/2}m_p)=8.46\cdot 10^{-5}$, where $G$ is the Newton constant and 
proton's charge-mass radio $|e|/(G^{1/2}m_p)=1.1\cdot 10^{18}$.

It is interesting to compare this rate of electron-positron pair-production with the rate given by the Hawking effect.
We take $R_c= 2GM/c^2$ and the charge-mass radio 
$Q/(G^{1/2}M)\simeq 10^{-19}$ for a naive balance between gravitational and electric forces. In this case 
$\beta = \frac{1}{2}(Q/G^{1/2}M)(|e|/G^{1/2}m)\approx 10^2$, 
the rate (\ref{rate2}) becomes,
\begin{eqnarray}
\Gamma_{\rm NS}
&=&  \frac{1.15}{6\pi^2}\left(\frac{25}{\tau\lambda^3}\right)\left(\frac{1}{mM}\right)\exp\left\{-0.492 (mM)\right\},
\label{rate24}
\end{eqnarray}
where the notation $mM=R_c/(2\lambda)$.
This is much larger than the rate of electron-positron emission by the Hawking effect \cite{hawking},
\begin{eqnarray}
\Gamma_{\rm H}
&\sim &  \exp\left\{-8\pi (mM)\right\},
\label{rateh}
\end{eqnarray}
since the exponential factor $\exp\left\{-0.492 (mM)\right\}$ is much larger 
than $\exp\left\{-8\pi (mM)\right\}$, where $2mM=R_c/\lambda\gg 1$.

\section{\it Summary and remarks}\label{conclusion}

In this letter, analogously to the study in atomic physics with large atomic number $Z$,
we study the classical and semi-classical (bound) states of electrons in the electric potential of a massive and charged 
core, which has a uniform charge distribution and macroscopic radius. We have found negative energy states of electrons inside the core, 
whose energies can be smaller than $-mc^2$, and the appearance of energy-level crossing to the negative energy spectrum. 
As a result, quantum tunneling takes place, leading to electron-positron pairs production, electrons then occupy these 
semi-classical (bound) states 
and positrons are repelled to infinity. Assuming that massive charged cores are bare and non of these semi-classical (bound)
states are occupied, we analytically obtain the maximal rate of electron-positron pair production in terms of the core
radius, charge and mass. We find that 
this rate is much larger than the rate of electron-positron pair-production by the Hawking effect, even for very small charge-mass radio of the core given by 
the naive balance between gravitational and electric forces.
\comment{
Any electron occupations of these semi-classical (bound) states must screen core's charge and the massive core is no longer bare. 
The electric potential potential inside the core is changed. For the core consists of a large number of electrons,  
the Thomas-Fermi approach has to be adopted. We recently study \cite{rrx2007,rrx2008} the electron distribution 
inside and outside the massive core, i.e., the distribution of electrons occupying stable states of the massive core, 
and find the electric field on the surface of the massive core is overcritical. 
}
%-------------------------1---------------------

%-------------------------2---------------------

%where $Ze^2$ is related to the charges of the core and electron, while $R_cm$ is related to
%the ratio between the core radius $R_c$ and the electron Compton wavelength $\lambda=\hbar/(mc)$.  
%We relate these parameters to the core charge-mass radio $\xi=Q/M$ and the electron charge-mass radio $e/m=2.04\cdot 10^{21}$,
%\begin{equation}
%\beta =1.02\cdot 10^{21}\frac{\xi}{R_c},
%\label{para}
%\end{equation}
%where $R_c$ in unit of $2M$. 

%From Eq.~(\ref{para}), we find for $R_c=10(2M)$
%\begin{equation}
%\beta=1.02\cdot 10^{21}\frac{\xi}{R_c}=1.02\cdot 10^{20}\xi >1,
%\label{para1}
%\end{equation}
%and the condition $\beta >\kappa\geq 1$ is satisfied, if $\xi > 1.0\cdot 10^{-20}$.


\begin{thebibliography}{99}

\bibitem {report}R. Ruffini, G. V. Vereshchagin, S.-S. Xue, Phys. Rep.,
Vol 487, (2010)  1.

\bibitem{z4}
P.~A.~M.~Dirac, Proc.~Roy.~Soc.~117 (1928) 610; P.~A.~M.~Dirac, Proc.~Roy.~Soc.~118 (1928) 341.

\bibitem{z5}
P.~A.~M.~Dirac, ``Principles of Quantum Mechanics'', Clarendon Press, Oxford, 1958.

\bibitem{z7}
W.~Gordon, Z.~Phys.~48 (1928) 11.

\bibitem{z6}
C.~G.~Darwin, Proc.~Roy.~Soc.~A118 (1928) 654.

\bibitem{Sommerfeld}
A.~ Sommerfeld, Atomau und Spektrallinien, 4.~ Aufl.~,6.~ Kap.

\bibitem{g1a}
I.~Pomeranchuk and J.~Smordinsky, J.~Phys.~USSR 9 (1945) 97.

\bibitem{g1b}
N.~Case, Phys.~Rev.~80 (1950) 797.

\bibitem{g1c}
F.~G.~Werner and J.~A.~Wheeler, Phys.~Rev.~109 (1958) 126.

\bibitem{g1d}
V.~Vorankov and N.~N.~Kolesinkov, Sov.~Phys.~JETP 12 (1961) 136.
%the first of reference of greiner's paper

\bibitem{z10}
V.~S.~Popov, Yad.~Fiz.~12 (1970) 429 [Sov.~J.~Nucl.~Phys.~12 (1971) 235].

\bibitem{z11}
V.~S.~Popov, Zhetf Pis.~Red.~11 (1970) 254 [JETP Lett.~11 (1970) 162]; V.~S.~Popov, Zh.~Eksp.~Theor Fiz.~59 (1970) 965 [Sov.~Phys.~JEPT 32 (1971) 526].

\bibitem{z12}
V.~S.~Popov, Zh.~Eksp.~Theor Fiz.~60 (1971) 1228 [Sov.~Phys.~JEPT 33 (1971) 665].

\bibitem{z}
Y.~B.~Zel'dovich and V.~S.~Popov, Sov.~Phys.~USPEKHI 14 (1972) 673.
%zeldovich's long paper,

\bibitem{popov1972}
M.~S.~Marinov and V.~S.~Popov, Pis'ma v ZhETF 17 (1973) 511 [JETP Lett.~17, (1973) 368 ];
G.~ Siff, B.~ M\"uller and J.~ Rafelski, Zeits.~ Naturforsch 292 (1974) 1267.

\bibitem{popov2001}
V.~S.~Popov, Yad.~Fiz.~ 64, (2001) 421 [Phys.~ Atomic Nuclei, 64 (2001) 367].

\bibitem{rfk78}
Review article: J.~Rafelski, L.P.~Fulcher and A.~Klein, Phys.~Reps.~38 (1978) 227.

\bibitem{gbook}
Review article: W.~Greiner, B.~M\"uller and J.~Rafelski, ``Quantum Electrodynamics of Strong Fields'' Monograph in Physics, ISBN 3-540-13404-2
Springer-Verlag Berlin Heidelberg (1985) (references there in).

\bibitem{krx2008}
H.~ Kleinert, R.~ Ruffini, S.-S.~ Xue, Phys.~ Rev.~ D78
(2008) 025001.

%\bibitem{z25}
%A.~B.~Migdal, D.~N.~Voskresenskii and V.~S.~Popov, Pis'ma Zh.~Eksp.~Tero.~Fiz.~24 (1976) 186 [Sov.~Phys.~JETP 24 (1976) 186].

\bibitem{sauter}
F.~Sauter, Z.~Phys.~69 (1931) 742.

\bibitem{euler}
W.~Heisenberg and H.~Euler, Z.~Phys.~98 (1936) 714.

\bibitem{schwinger}
J.~Schwinger, Phys.~Rev.~82 (1951) 664; J.~Schwinger, Phys.~Rev.~93 (1954) 615; J.~Schwinger, Phys.~Rev.~94 (1954) 1362.
\comment{
\bibitem{rrx2007}
R.~Ruffini, M.~Rotondo and S.-S.~Xue, Int.~ Journal of Modern Phys. D  Vol.~16, No.~1 (2007) 1-9, astro-ph/0609190.
\bibitem{rrx2008}
M.~Rotondo, R.~Ruffini and S.-S.~Xue, to be submitted to Phys.~Rev.~Lett.
}

\bibitem{hawking}
S.~W.~Hawking, Nature 238 (1974) 30; S.~W.~Hawking, Commun.~Math.~Phys.~43, 199 (1975); G.~W.~Gibbons and S.~W.~Hawking, Phys.~Rev.~D 15 2752 (1977).


\end{thebibliography}
\end{document}